\documentclass[useAMS,useamsbsy,usetimes,usemathptm,usenatbib,usegraphicx]{mn2e}

\title[Chaos in the problem of jet collimation]{Dynamical chaos in the problem of magnetic jet collimation}
\author[G. S. Bisnovatyi-Kogan et al]{G. S. Bisnovatyi-Kogan$^{1,2}$\thanks{E-mail:
gkogan@iki.rssi.ru (GSBK); aneishta@iki.rssi.ru (AIN); seidov@bgu.ac.il (ZFS); tsupko@iki.rssi.ru (OYuT);
krivosheev@iki.rssi.ru (YuMK)}, A. I. Neishtadt$^{1,3}$\footnotemark[1], Z. F. Seidov$^{4}$\footnotemark[1],
\newauthor O. Yu. Tsupko$^{1,2}$\footnotemark[1] and Yu. M. Krivosheyev$^{1}$\footnotemark[1] \\
$^{1}$Space Research Institute of Russian Academy of Sciences, Profsoyuznaya 84/32, Moscow 117997, Russia\\
$^{2}$National Research Nuclear University MEPhI, Kashirskoe Shosse 31, Moscow 115409, Russia\\
$^{3}$Department of Mathematical Sciences, Loughborough University, Loughborough, LE11 3TU, UK\\
$^{4}$Ben Gurion University, Beer-Sheva, 84105, Israel}
\begin{document}



\maketitle

\label{firstpage}

\begin{abstract}
We investigate  dynamics of a jet collimated by magneto-torsional oscillations. The problem is reduced to an
ordinary differential equation containing a singularity and depending on a parameter. We find a parameter
range for which this system has stable periodic solutions and study bifurcations of these solutions. We use
Poincar\'e sections to demonstrate existence of domains of regular and chaotic motions. We investigate
transition from periodic to chaotic solutions through a sequence of period doublings.
\end{abstract}


\section{Introduction}

\begin{figure}
\centerline{\hbox{\includegraphics[width=0.5\textwidth]{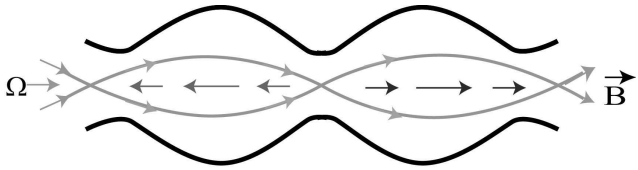}}} \caption{Jet confinement by magneto-torsional
oscillations (qualitative picture).} \label{jet2bw}
\end{figure}

Many quasars and active galactic nuclei are connected with long thin collimated outbursts -- jets. When
observed with high angular resolution, these jets show structure with bright knots separated by relatively
dark regions. A mechanism of collimation of such jets is still questionable. Magnetic collimation of jets was
first considered by \citet*{bkf69}. In the paper of \citet{bk07} magnetic collimation resulting from
torsional oscillations of a cylinder with elongated magnetic field and periodically distributed initial
rotation around the cylinder axis was considered (Fig. \ref{jet2bw}). The stabilizing azimuthal magnetic
field is created here by torsional oscillations. An approximate simplified model was developed, and an
ordinary differential equation was derived describing the process of dynamic stabilization. The interval of
parameters, for jet stabilization to occur, was estimated qualitatively.

The ordinary differential equation under consideration is a non-linear non-autonomous time-periodic second
order equation with a singularity in the right hand side. The equation contains one dimensionless parameter
$D$, which summarizes the information about the magnetic field, amplitude and frequency of oscillations,
radius of the jet, its spatial period along the jet axis, and sound speed in the jet matter. Here we
investigate analytically and numerically the structure of the phase space of this equation, which has a very
peculiar character and contains chaotic solutions as well as quasi-periodic and periodic regular solutions.

The paper is organized as follows. In section 2 we describe in more detail the mechanism of jet collimation
by magneto-torsional oscillations and the dynamic confinement equation. In section 3 we discuss the main
types of solutions of this equation. In section 4 we investigate analytically and numerically the dynamics of
the system for large radii. In section 5 we construct Poincar\'e sections for different values of parameter
$D$. In section 6 we study periodic solutions undergoing a sequence of period doublings at values of
parameter $D=D_n$, $n=$1,2,3,\ldots, when $D$ increases. In section 7 we discuss a possible mechanism of jet
formation with a variable direction of angular velocity and estimate the value of $D$ for astrophysical jets.

We have found that stable periodic solutions disappear after an infinite cascade of period doublings. In the
problem of period doubling the limiting constant $q$ for values $[D_n-D_{n-1}]/[D_{n+1}-D_n]$ is usually
considered. For large $n$ the constant $q$ equals to Feigenbaum constant $F=4.6692...$ for dissipative
systems, see \cite{feig80}, and $FH=8.721...$ for Hamiltonian systems, see \cite{feig81}. In our work we
obtain that $q$ is approaching 8.72, in agreement with the expected behaviour for Hamiltonian systems.

\section{Dynamic confinement of jets by magnetotorsional oscillations}

We consider jet stabilization by pure magneto-hydrodynamic mechanism associated with torsional oscillations.
Jets are relativistically moving objects, and the whole analysis of present paper takes place at the rest
frame of the bulk motion of the jet. We suggest that matter in the jet is rotating, and different parts of
the jet rotate in different directions (see Fig. \ref{jet2bw}). Such distribution of rotational velocity
produces azimuthal magnetic field, which prevents the disruption of the jet. The jet is represented by a
periodic, or quasi-periodic structure along the axis, and its radius oscillates with time all along the axis.
Space and time periods of oscillations depend on the conditions of jet formation: the length-scale, the
amplitude of the rotational velocity, and the strength of the magnetic field. Time period of oscillations can
be calculated in the framework of the dynamical model. The range of parameters, for which dynamical
stabilization occurs, should also be inferrible from the model. Two-dimensional non-stationary MHD
calculations are needed to solve the problem numerically. A very simplified model of this phenomenon was
constructed by \citet{bk07}. This model allowed to confirm the possibility of such stabilization, to estimate
the interval of parameters when it takes place and to establish the connection between time and space scales,
magnetic field strength, and amplitude of rotational velocity.

Let us consider a long cylinder with magnetic field directed along its axis. This cylinder will expand
without limit under the action of pressure and magnetic forces. It is possible, however, that a limiting
value of the cylinder radius could be reached in a dynamic state, when the whole cylinder undergoes
magneto-torsional oscillations. Such oscillations produce a toroidal field (magnetic field lines are frozen
into matter), which prevents radial expansion. There is therefore a competition between the induced toroidal
field, compressing the cylinder in the radial direction, and the gas pressure, together with the field along
the cylinder axis (poloidal), tending to increase its radius. During magneto-torsional oscillations there are
phases when either compression or expansion force prevails, and, depending on the input parameters, there are
three possible kinds of behaviour of such a cylinder that has negligible self-gravity.

(1) The oscillation amplitude is low, so the cylinder suffers unlimited expansion (no confinement).

(2) The oscillation amplitude is high, so the pinch action of the toroidal field destroys the cylinder and leads to the
formation of separated blobs.

(3) The oscillation amplitude is moderate, so the cylinder, in absence of any damping,  survives for an
unlimited time, and its parameters (radius, density, magnetic field, etc.) change periodically, or
quasi-periodically, or chaotically in time.

These phenomena can be described in general by the system of axially symmetric MHD equations \citep{bk07}.
The system of equations was simplified in the paper of \citet{bk07} to investigate the most important
property of dynamical competition between different forces in order to check for the possibility of dynamical
confinement. A profiling procedure was used for this purpose. In particular, gravity in the direction of the
cylinder axis was neglected, approximate uniform density along the radius was assumed, and adiabatic case
with polytropic equation of state was considered. Approximate system allowed to investigate linear
oscillations (around the equilibrium state in presence of gravity) of infinite, self-gravitating cylinder
with uniform magnetic field and rotation along its axis.

For further simplification gravity was neglected, because in a relativistic jet the self-gravitating force is
expected to be much lower than the magnetic and pressure forces. In this approximation we have the following
feature of problem. Without gravity the equilibrium static state of the cylinder does not exist. But it turns
out that nevertheless bounded solutions exist, where cylinder radius oscillates and remains finite (dynamic
confinement). Confinement by magneto-torsional oscillations is therefore physically realizable. Similar
situation is observed for instance in the simulations of electron-positron discharges in the polar cap of
pulsar -- the pair cascade settles down to a stable state where it has a limiting cycle type of behaviour
\citep{Timokhin}.

After considerable simplifications, details of which may be found in the paper of \cite{bk07}, the equation,
describing the magneto-torsional oscillations of a long cylinder, takes the following form:

\begin{equation}
 \label{eq1}
 \frac{d^2\,y}{d\,\tau^2}=\frac{1-D\,\sin^2\,\tau}{y}.
\end{equation}
This equation describes approximately time dependence of the outer radius of the cylinder $R(t)$ in the
symmetry plane, where the rotational velocity remains zero. The cylinder has magnetic field $B_z(t)$,
isothermal equation of state of matter $P=K\rho$, maximal amplitude of the angular velocity of oscillations
$\Omega_0$, and angular frequency of oscillations $\omega$. The oscillating cylinder has a periodic structure
along the $z$ axis, with space period $z_0$. The nodes with zero amplitude of oscillations  are situated at
$z=\pm n\frac{z_0}{2}$, and the oscillations with maximal angular amplitude $\Omega_0$ are situated at $z=
\frac{z_0}{4} \pm n\frac{z_0}{2}$, $n=0,1,2 ...$. Axial motion of the matter in the cylinder is neglected,
and density $\rho(t)$ is supposed to be uniform along the radius. Under these conditions the equations for
conservation of mass and magnetic flux (we assume infinite electrical conductivity) determine the constant
values $C_m$ and $C_b$:

\begin{equation}
\label{eq2} \rho\, R^2 = C_m=\rho_0 R_0^2,\quad B_z\, R^2 = C_b=B_{z,0}R_0^2,
\end{equation}
where $\rho_0$, $R_0$, and $B_{z,0}$ are some characteristic values. The dimensionless variables and the
parameter $D$ in (\ref{eq1}) are defined as

\begin{equation}
\label{eq3} \tau=\omega \, t, \quad y=\frac{R}{R_0},
\end{equation}
\[
\hbox {with} \quad R_0=\frac{\sqrt{K}}{\omega}, \quad D=\frac{1}{2\pi K C_m}\left(\frac{C_b
\Omega_0}{z_0\omega}\right)^2.
\]

The frequency of oscillations $\omega$ may be represented as

\begin{equation}
 \label{eq4}
 \omega=\alpha_n k V_A=\alpha_n\frac{B_{z,0}}{z_0}\sqrt{\frac{\pi}{\rho_0}},
\end{equation}
where $k$ is the wave number,  $k=2\pi/z_0$, and $V_A$ is the Alfven velocity,
$V_A=B_{z,0}/\sqrt{4\pi\rho_0}$; $\alpha_n<1$ is a coefficient determining the frequency of non-linear Alfven
oscillations, which are identical to magneto-torsional oscillations under investigation. In our problem there
are no static solutions. In a balanced non-compressible cylinder the frequency of magneto-torsional
oscillations is defined by (\ref{eq4}) with $\alpha_n=1$.

\section{General types of solutions of dynamic confinement equation}

Equation (\ref{eq1}) was solved numerically in the paper of \cite{bk07} for different values of parameter
$D$, and fixed initial conditions $y(0)=1,  y'(0)=0$. The following three regimes of the behaviour were found
in these calculations.

\begin{figure}
\centerline{\hbox{\includegraphics[width=0.5\textwidth]{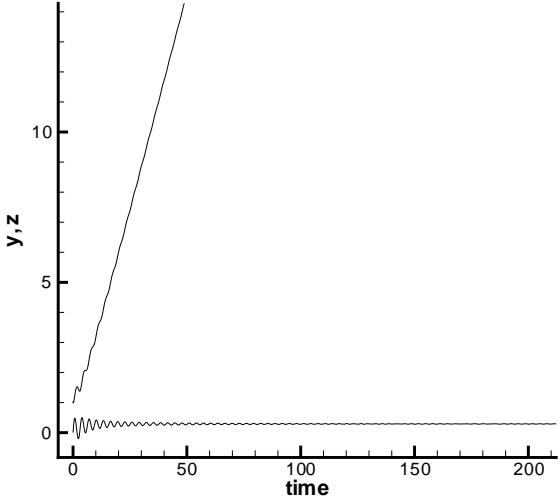}}} \caption{Time dependence of
non-dimensional radius $y$ (upper curve), and non-dimensional velocity $z=y'$ (lower curve), for $D=2.0$.}
\label{bk2-0}
\end{figure}

(1) $D<2.1$ The oscillation amplitude is low, so the cylinder should suffer unlimited expansion (no
confinement). This regime corresponds to small $D$. For example at  $D=2$ there is no confinement, and radius
grows to infinity after several low-amplitude oscillations, Fig. \ref{bk2-0}.

\begin{figure}
\includegraphics[width=0.5\textwidth]{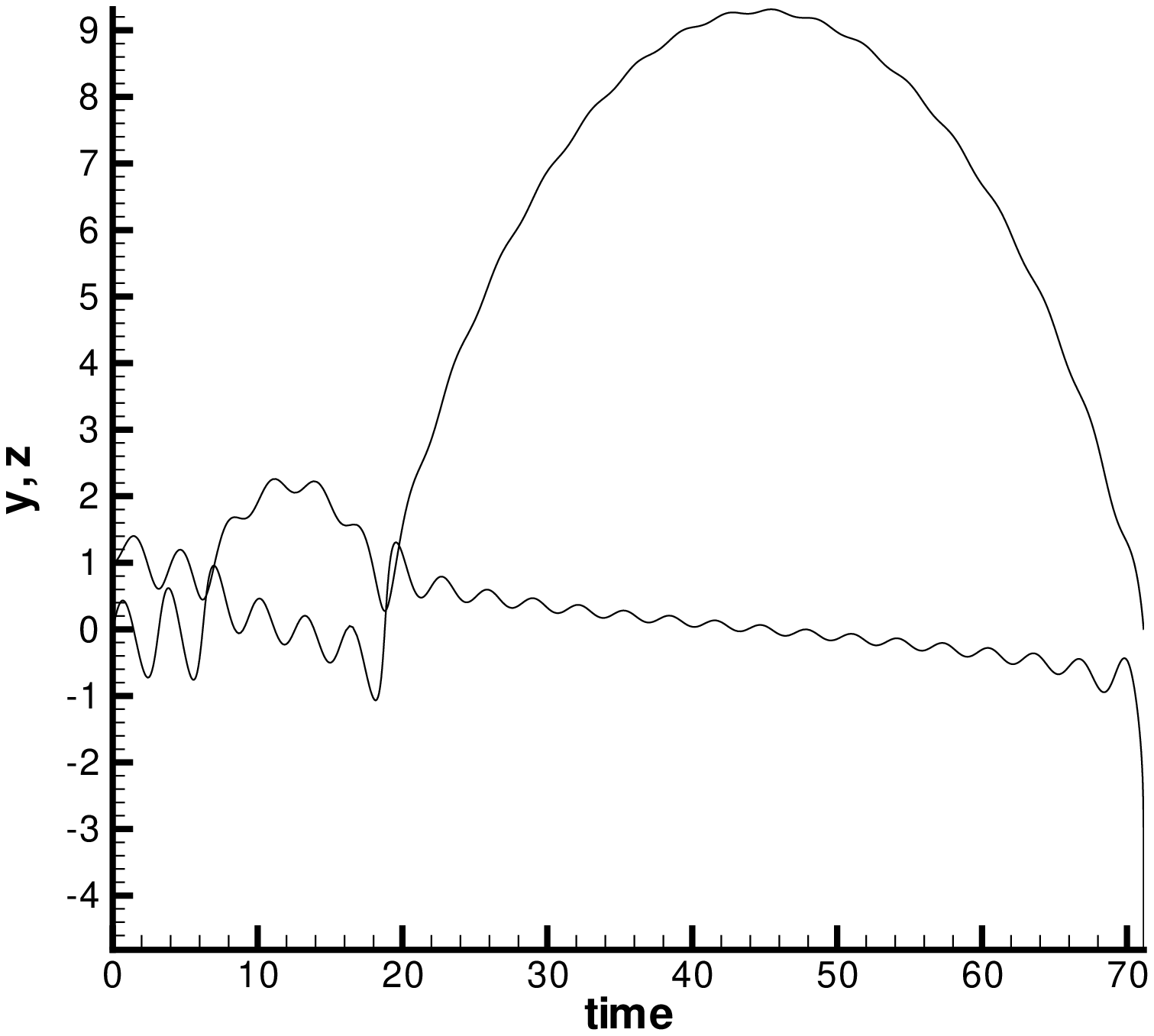}
\caption{Time dependence of non-dimensional radius $y$ (upper curve), and non-dimensional velocity $z=y'$ (lower
curve), for $D=2.4$.} \label{bk2-4}
\end{figure}

\begin{figure}
\centerline{\hbox{\includegraphics[width=0.5\textwidth]{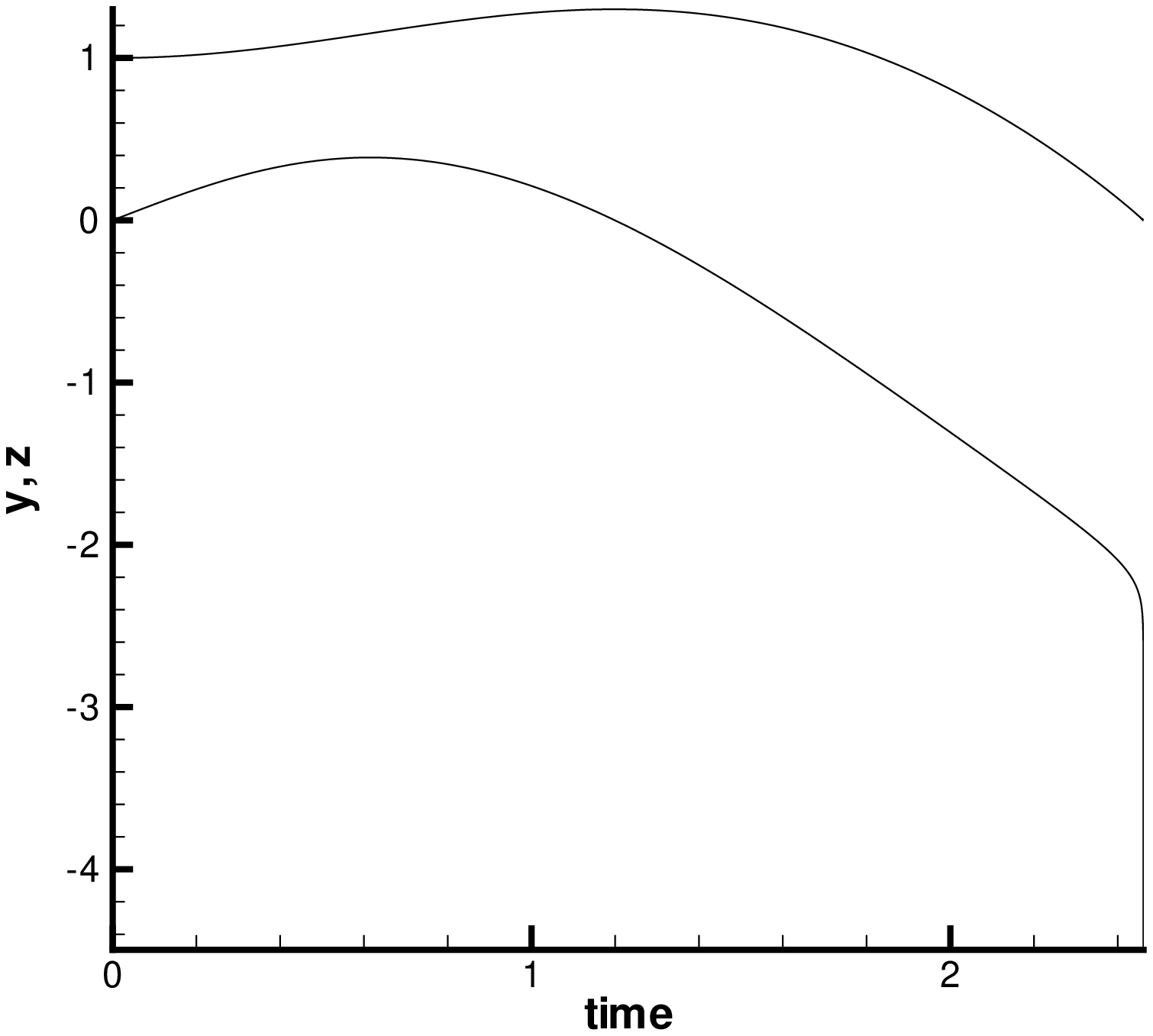}}} \caption{Time dependence of
non-dimensional radius $y$ (upper curve), and non-dimensional velocity $z=y'$ (lower curve), for $D=3.0$.}
\label{bk3-0}
\end{figure}

\begin{figure}
\centerline{\hbox{\includegraphics[width=0.5\textwidth]{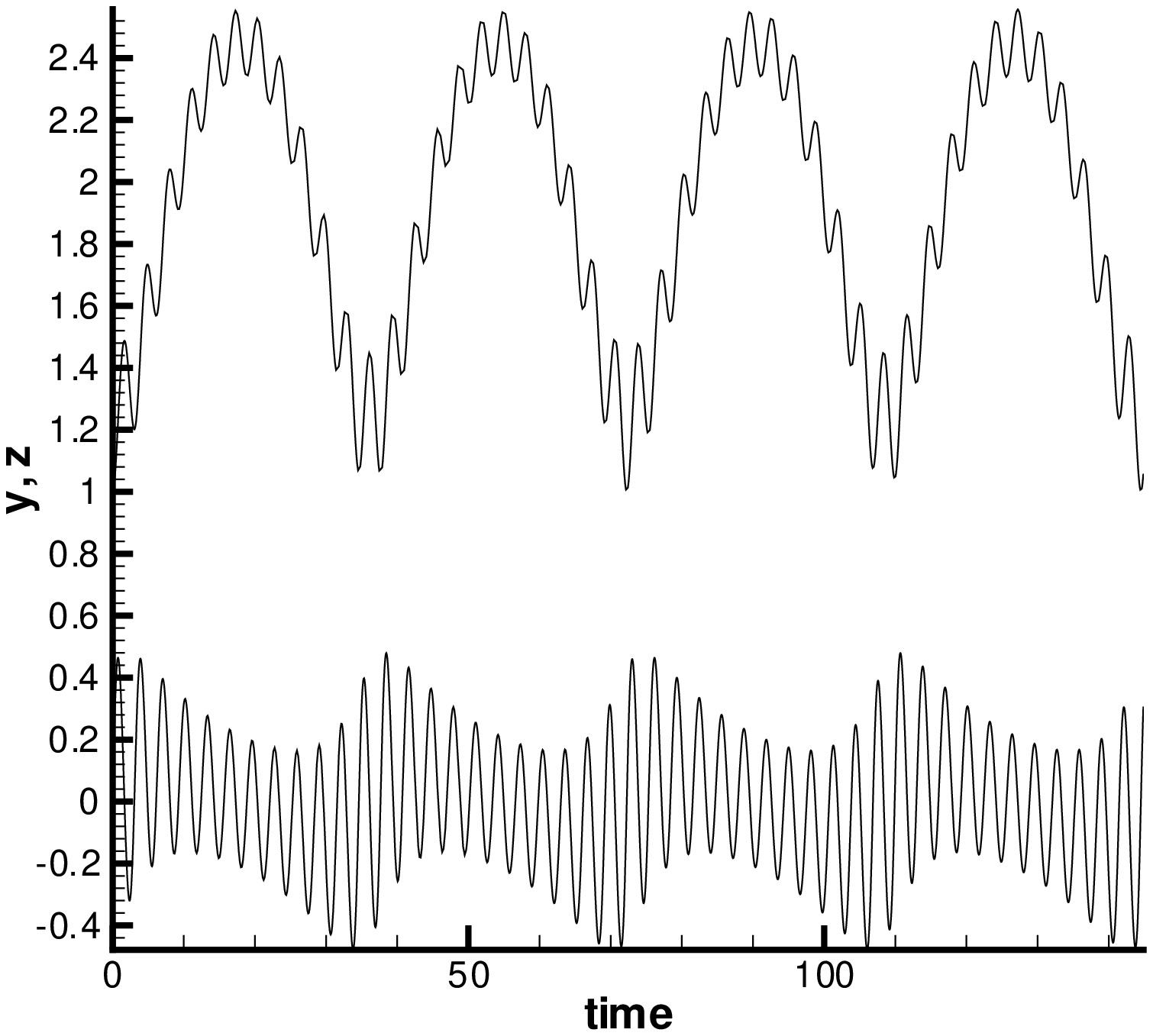}}} \caption{Time dependence of
non-dimensional radius $y$ (upper curve), and non-dimensional velocity $z=y'$ (lower curve), for $D=2.1$.}
\label{bk2-1}
\end{figure}

\begin{figure}
\centerline{\hbox{\includegraphics[width=0.5\textwidth]{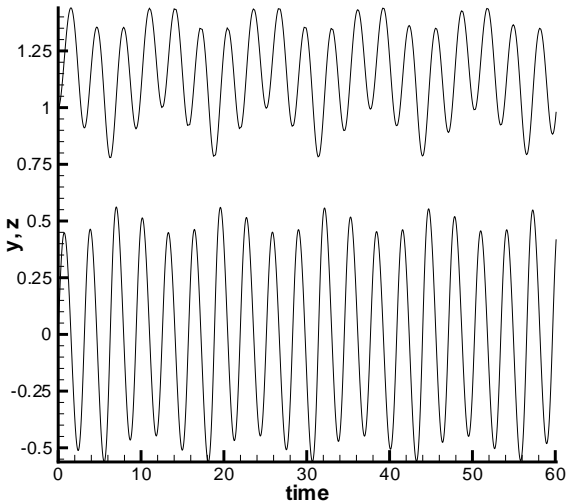}}} \caption{Time dependence of
non-dimensional radius $y$ (upper curve), and non-dimensional velocity $z=y'$ (lower curve), for $D=2.25$.}
\label{bk2-25}
\end{figure}

(2) $D>2.28$ The oscillation amplitude is too high, so the pinch action of the toroidal field destroys the
cylinder, and leads to formation of separated blobs. The calculations show \citep{bk07}, that at $D= 2.28$
and larger, the radius finally goes to zero with time, but with different behaviour, depending on $D$. At $D$
between 2.28 and 2.9 time dependence of the radius $y$  may be very complicated, consisting of low-amplitude
and large-amplitude oscillations, which finally decay  to zero. The time at which radius becomes  zero
depends on $D$ in a rather peculiar way, and it may happen at $\tau<100$, like at $D=2.4$ (Fig. \ref{bk2-4}),
$2.6$, or at $\tau\sim 10^7$ like  at $D=2.5$ (in the last case the radius passes through very large values
and  then returns back to zero). For  $D= 3$ and larger the solution is very simple: the radius goes to zero
at $\tau < 2.5$, before the return of the right hand side of (\ref{eq1}) to a positive value, Fig.
\ref{bk3-0}.

(3) For the intermediate interval  $2.1< D<2.28$ the radius is not growing to infinity, but is oscillating
around some average value. The cylinder survives for a unlimited time, and its parameters (radius, density,
magnetic field, etc.) change periodically or quasi-periodically in time, Figs \ref{bk2-1}, \ref{bk2-25}.

The above results from the paper of \citet{bk07} have been obtained for solutions only with initial radius $y_0=1$. The
solutions at moderate $D$ are not pure periodic, but it is not clear from the figures, whether are they regular or
chaotic.

In the current paper  we investigate the solutions of (\ref{eq1}) for different values of $D$ and different
initial radii $y_0$, mainly for moderate $D$'s. Solutions of principal interest are those that do not go
neither to 0 nor to infinity. Such solutions correspond to stabilized jets. We use averaging method and
Poincar\'e sections to investigate the properties of these solutions.

\section{Dynamics for large radii}

\begin{figure}
\centerline{\hbox{\includegraphics[width=0.45\textwidth]{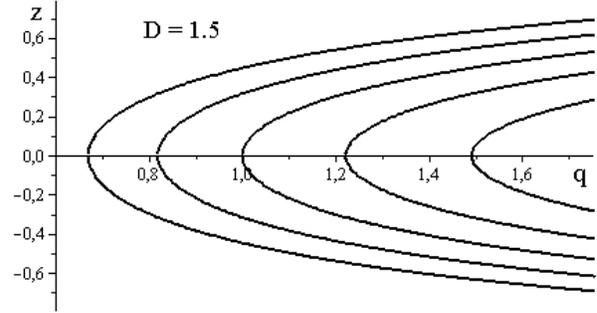}}} \caption{Phase curves of the system
with Hamiltonian (\ref{H_av}) for $D=1.5$, and ${\cal H} = -0.1, -0.05, 0 , 0.05, 0.1$.} \label{phase1-5}
\end{figure}

\begin{figure}
\centerline{\hbox{\includegraphics[width=0.45\textwidth]{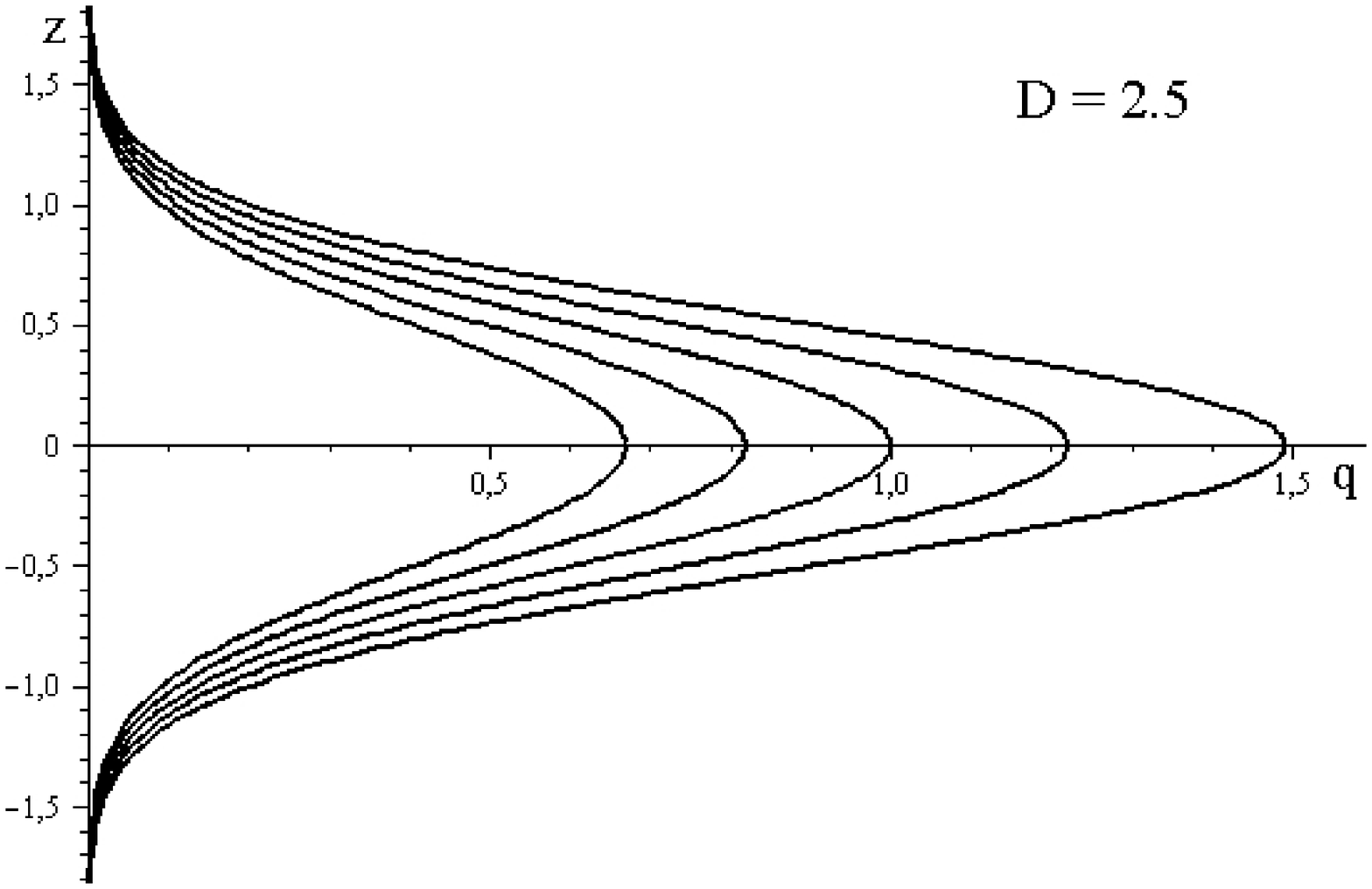}}} \caption{Phase curves of the system
with Hamiltonian (\ref{H_av}) for $D=2.5$, and ${\cal H} = -0.1, -0.05, 0 , 0.05, 0.1$.} \label{phase2-5}
\end{figure}

Let us denote $z= y' \equiv dy/d \tau$. Equation  (\ref{eq1}) is equivalent to the system

\begin{equation}
\label{syst} y'=z, \quad z'=\frac{1-D\,\sin^2\,\tau}{y}.
\end{equation}
We consider a motion for large values of $y$. In this limit let us introduce a small number $\varepsilon$ and
denote $q=\varepsilon y$. We will assume that $y\sim 1/\varepsilon$, and therefore $q\sim 1$. The equation
system for $q,\, z$ is

\begin{equation}
q'=\varepsilon z, \quad z'=\varepsilon \frac{1-D\,\sin^2\,\tau}{q}.
\end{equation}
This is a Hamiltonian system with Hamiltonian

\begin{eqnarray}
\varepsilon H=\varepsilon\left [\frac{z^2}{2}-(1-D\sin^2\tau)\ln q\right ]\label{H_nonav}
\end{eqnarray}
and equations of motion

\begin{eqnarray}
q'=\varepsilon {\partial H}/{\partial z},\quad z'=-\varepsilon{\partial H}/{\partial q}.
\end{eqnarray}
For small $\varepsilon$ the variables $q,z$ are changing slowly with respect to changing of $\tau$. The
averaging method \citep{bm} prescribes to average Hamiltonian (\ref{H_nonav}) over the fast variable $\tau$
for an approximate description of the behaviour of the slow variables $q,z$. We get the averaged Hamiltonian

\begin{equation}
\varepsilon {\cal H}=\varepsilon\left [\frac{z^2}{2}-\left(1-\frac{D}{2}\right)\ln q\right ].\label{H_av}\\
\end{equation}
Phase curves of the system with Hamiltonian (\ref{H_av}) for $D<2$, and for $D>2$ are shown in Figs
\ref{phase1-5} and \ref{phase2-5}, respectively.

If $D<2$, then for Hamiltonian (\ref{H_av}) $q(t)$ and $y(t)\to  \pm \infty, z(t)\to \pm \infty$ as $t\to
\pm\infty$ for all solutions of the averaged system. This situation corresponds to the unlimited expansion of
the jet (no collimation). If  $D>2$, then $q(t)$ and $y(t) \to 0, z(t)\to \pm \infty$ as $t\to \mp \tau_*$,
where $\tau_*$ is some finite moment of time that depends on  the chosen solution. This corresponds to strong
initial amplitude of pulsations leading to the fragmentation of the jet into separate clumps.

\begin{figure}
\centerline{\hbox{\includegraphics[width=0.4\textwidth]{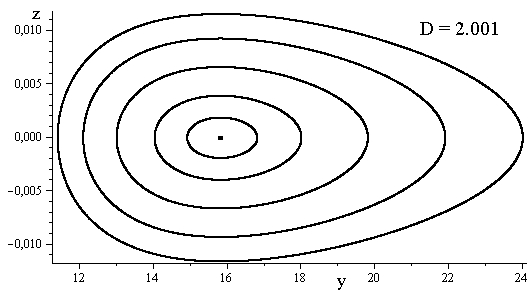}}} \caption{Poincar\'e section near the
stable fixed point, $z= y' \equiv dy/d \tau$, $D=2.001$. One periodic and five quasi-periodic solutions are
shown. This figure is constructed numerically by solutions of equations (\ref{syst}). Difference between
curves in this figure and phase curves from approximate formula (\ref{port}) is about thickness of line on
figure, therefore we do not show the latter curves here.} \label{2-001}
\end{figure}

\begin{figure}
\centerline{\hbox{\includegraphics[width=0.4\textwidth]{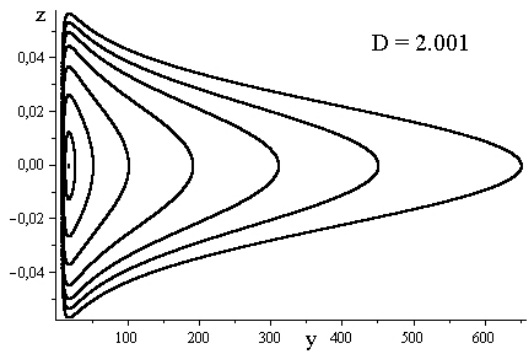}}} \caption{Poincar\'e section for
$D=2.001$. This figure is constructed numerically by solutions of equations (\ref{syst}). Difference between
curves in this figure and phase curves from approximate formula (\ref{port}) is about thickness of line on
figure, therefore we do not show the latter curves here.} \label{2-001big}
\end{figure}

Solutions of the averaged system describe approximately the solutions of the exact system for large values of
$y$ provided that $(D-2)$ is not small.

It turns out that for small $(D-2)>0$ the exact system has a stable periodic solution, which is described
approximately by the formulas

\begin{equation}
y=\frac{1}{2\sqrt{D-2}}-\frac{1}{2}\sqrt{D-2}\cos2\tau, \quad z=\sqrt{D-2}\sin 2\tau.
\end{equation}
(see Appendix). In the Poincar\'e section $\{\tau = 0\, {\rm mod} \,\,\pi\,\}$ this solution is depicted as a
fixed point of the Poincar\'e return map. This fixed point is located on the axis $z=0$, its $y$-coordinate
is  given approximately by the formula

\begin{equation}
y=\frac{1}{2\sqrt{D-2}}-\frac{1}{2}\sqrt{D-2}. \label{t_y}
\end{equation}
This fixed point is surrounded by  a family of invariant curves of the Poincar\'e return map. The approximate
equation for these invariant curves is

\begin{equation}
\frac{1}{2}z^2+ \frac{1}{16y^2}+\frac{1}{2}(D-2)\ln y = {\rm const}
\label{port}
\end{equation}
(see Appendix). The fixed point and surrounding it invariant curves are shown in Fig. \ref{2-001} for $D=2.001$.

The Poincar\'e sections are constructed numerically in this paper. For the same value of D, we solve
equations (\ref{syst}) for different initial $y_0$ and $z_0$. During the numerical integration of equations,
we mark moments $\tau = 0\,\, {\rm mod}\,\, \pi$. For each integration, we put the points on the plane $(y,
z)$ at the moments $\tau = 0\,\, {\rm mod}\,\, \pi$. The regular oscillations are represented by closed lines
(invariant curves) on the Poincar\'e maps: closed curves correspond to regular quasi-periodic solutions,
points in the centers of families of such curves correspond to stable regular periodic solutions. Chaotic
behaviour fills regions of finite area with dots on the Poincar\'e maps.

\begin{figure}
\centerline{\hbox{\includegraphics[width=0.45\textwidth]{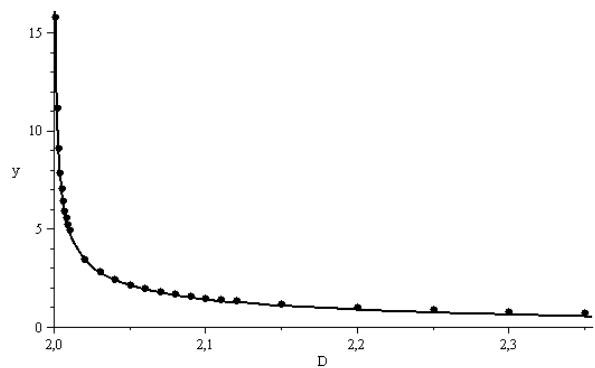}}} \caption{$y$-coordinate of the stable
fixed point as a function of $D$. In this figure the separated points are obtained by numerical calculations,
the curve is plotted analytically with formula (\ref{t_y}).} \label{th-num}
\end{figure}

\begin{figure}
\centerline{\hbox{\includegraphics[width=0.4\textwidth]{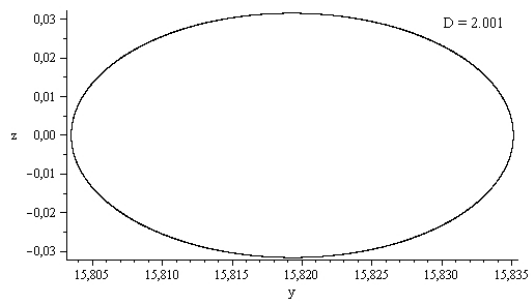}}} \caption{Phase trajectory
corresponding to periodic solution for $D=2.001$; initial data $y(0)=15.8034860625$, $z(0)=0$. Phase
trajectory is constructed by numerical solution of equations (\ref{syst}) at given initial values and putting
all dots of solution on the plane $(y,z)$.} \label{2-001fd}
\end{figure}

For this value of $D=2.001$ the  family of invariant curves covers rather a big domain in the Poincar\'e
section, see Fig. \ref{2-001big}. For initial conditions on the invariant curves the solution of the equation
(\ref{eq1}) is represented by quasi-periodic functions of $\tau$. The values of $y$-coordinate of the fixed
point, obtained theoretically (see (\ref{t_y})) and numerically, are shown in Fig. \ref{th-num}.

One can see that the obtained asymptotic expression for the coordinates (radii) of the stable fixed points
works very well even for $D$ not very close to 2 (and therefore when $y$ is not very big). The projection
onto the plane $y, z$ (in another words, phase trajectory) of the periodic solution corresponding to the
fixed point in Fig. \ref{2-001} is shown in Fig. \ref{2-001fd}.

\section{Poincar\'e sections for different values of parameter $D$}

In order to find bounded regular solutions we have constructed a Poincar\'e section for the system
(\ref{syst}) for $\tau = 0\,\, {\rm  mod}\,\, \pi$ for different values of parameter $D$. We use such
construction of the Poincar\'e section because the right-hand side of the equation (\ref{eq1}) is a periodic
function of $\tau$ with period $\pi$. Recall that in \cite{bk07} such solutions with $y_0=1$ were found to
exist only for $\sim 2.1<D<\sim 2.28$. We will concentrate here on the behaviour of the invariant curves on
the Poincar\'e section around the stable fixed point of the Poincar\'e return map, and on the transition to
chaotic behaviour in this region with increasing of parameter $D$.

\begin{figure}
\centerline{\hbox{\includegraphics[width=0.465\textwidth]{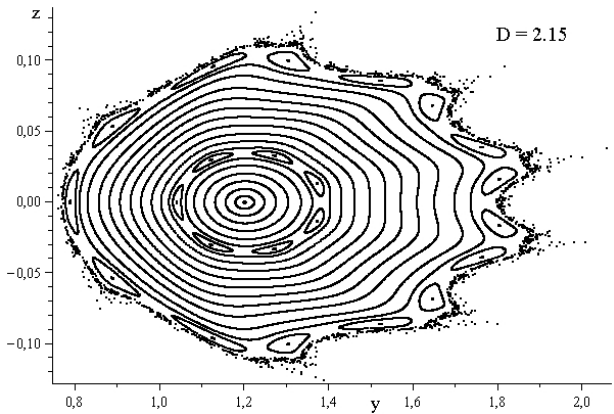}}} \caption{Poincar\'e section for
$D=2.15$. Central point corresponds to a stable periodic solution of period $\pi$.} \label{2-15}
\end{figure}

\begin{figure}
\centerline{\hbox{\includegraphics[width=0.35\textwidth]{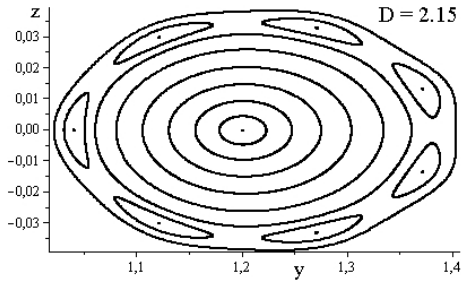}}} \caption{Zoom of the part of Fig.
\ref{2-15}: islands on Poincar\'e section for $D=2.15$.} \label{2-15big}
\end{figure}

In Figs \ref{2-15}, \ref{2-15big} the Poincar\'e section for $D=2.15$ is presented. The closed invariant
curves cross axis $z=0$ in the interval $0.8 \sim<y_0< \sim 1.80$. Outside this interval solutions of
equations (\ref{syst}) show non-regular chaotic behaviour, with unbounded trajectories. Central point in Fig.
\ref{2-15} corresponds to the periodic solution with period $\pi$. One can see a chain of seven stability
islands, centres of which determine periodic solution with period $7 \pi$; zoom of this chain is presented in
Fig. \ref{2-15big}. In Fig. \ref{2-15} one can also see a chain of fifteen stability islands, centres of
these islands correspond to periodic solution with period $15 \pi$.

\begin{figure}
\centerline{\hbox{\includegraphics[width=0.35\textwidth]{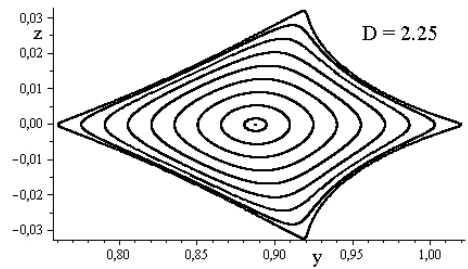}}} \caption{Invariant curves on Poincar\'e
section for $D=2.25$. The stable fixed point in the centre has coordinates $y=0.887$, $z=0$.} \label{2-25}
\end{figure}

\begin{figure}
\centerline{\hbox{\includegraphics[width=0.4\textwidth]{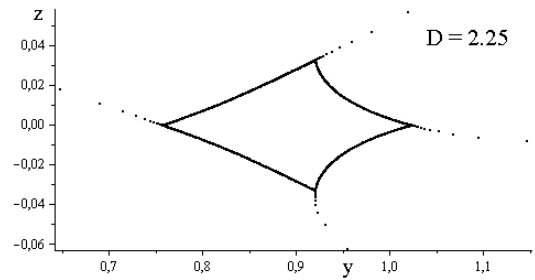}}} \caption{Appearance of chaos on
Poincar\'e section for $D=2.25$.} \label{2-25add}
\end{figure}

\begin{figure}
\centerline{\hbox{\includegraphics[width=0.45\textwidth]{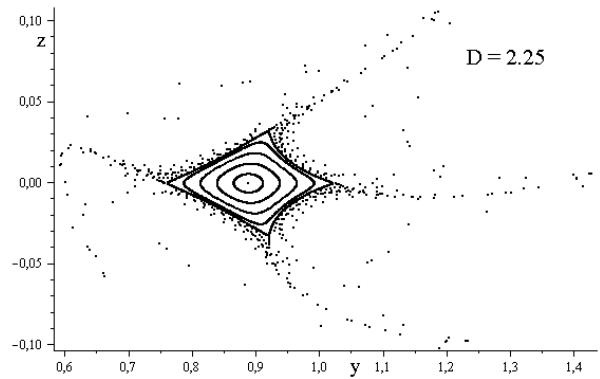}}} \caption{Invariant curves and
chaotic trajectories on Poincar\'e section for $D=2.25$.} \label{2-25add2}
\end{figure}

Invariant curves around the periodic solution for $D=2.25$ are plotted in Fig. \ref{2-25}. In Fig.
\ref{2-25add} the appearance of a chaotic solution is shown. The last shown invariant curve corresponds to
the solution with initial values $y(0)=0.757$, $z(0)=0$, while the solution with $y(0)=0.756$, $z(0)=0$ is
already chaotic. In Fig. \ref{2-25add2} a general picture of regular and  chaotic trajectories is
represented.

\begin{figure}
\centerline{\hbox{\includegraphics[width=0.45\textwidth]{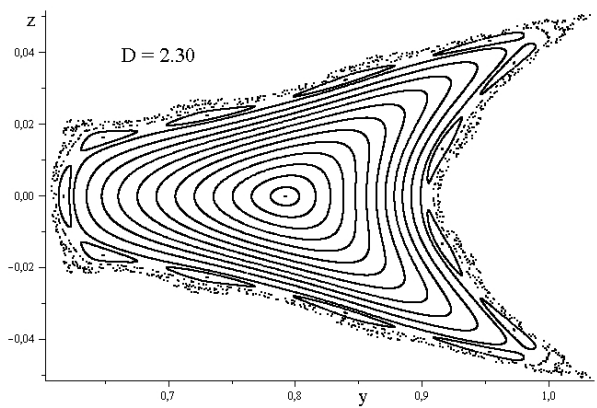}}} \caption{Invariant curves and chaotic
trajectories on Poincar\'e section for $D=2.30$. Central point corresponds to a stable periodic solution of
period $\pi$.} \label{2-30}
\end{figure}

\begin{figure}
\centerline{\hbox{\includegraphics[width=0.45\textwidth]{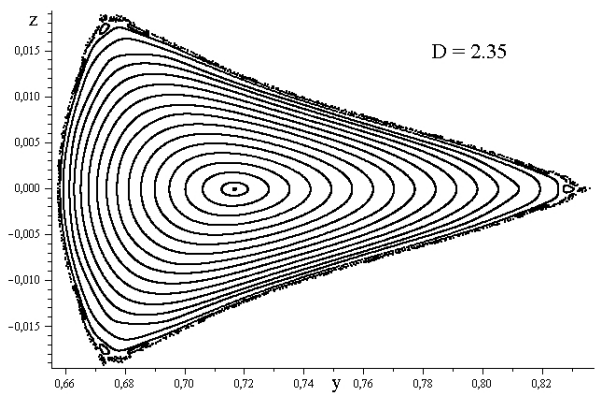}}} \caption{Invariant curves and chaotic
trajectories on Poincar\'e section for $D=2.35$. Central point corresponds to a stable periodic solution of
period $\pi$.} \label{2-35}
\end{figure}

Invariant curves and chaotic trajectories on the Poincar\'e section are plotted for $D=2.30$ in Fig.
\ref{2-30}, and for $D=2.35$ in Fig. \ref{2-35}.

\begin{figure}
\centerline{\hbox{\includegraphics[width=0.45\textwidth]{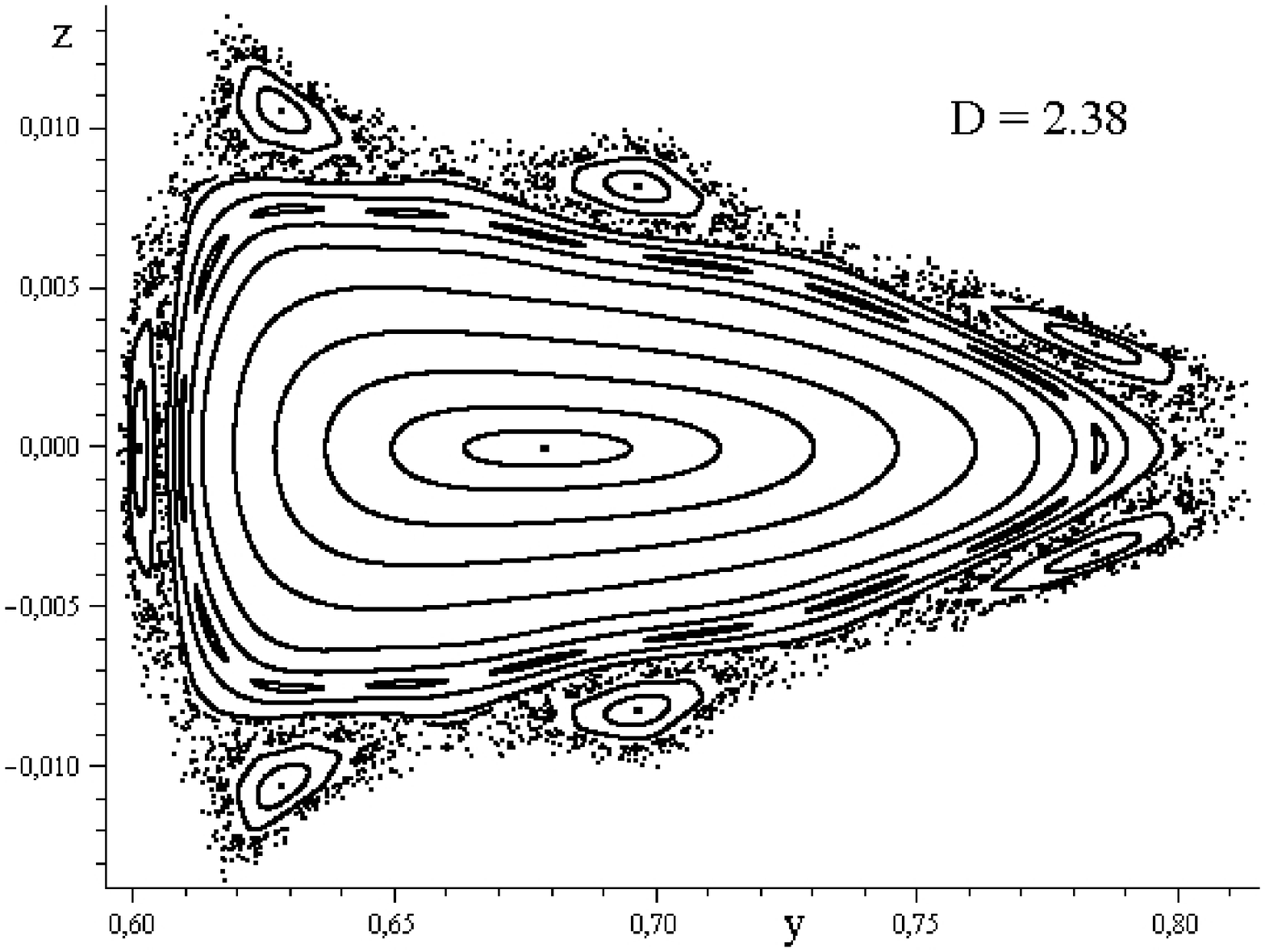}}} \caption{Invariant curves and chaotic
trajectories on Poincar\'e section for $D=2.38$. Central point corresponds to a stable periodic solution of
period $\pi$.} \label{2-38}
\end{figure}

\begin{figure}
\centerline{\hbox{\includegraphics[width=0.45\textwidth]{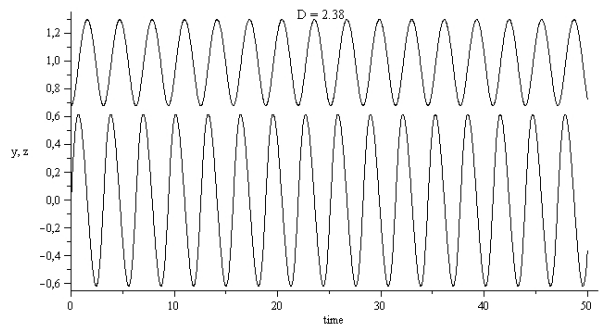}}} \caption{Time dependence of
non-dimensional radius $y$ (upper curve) and non-dimensional velocity $z = y'$ (lower curve) for periodic
solution of period $\pi$, see central point in Fig. \ref{2-38}; $D=2.38$.} \label{2-38ev}
\end{figure}

\begin{figure}
\centerline{\hbox{\includegraphics[width=0.45\textwidth]{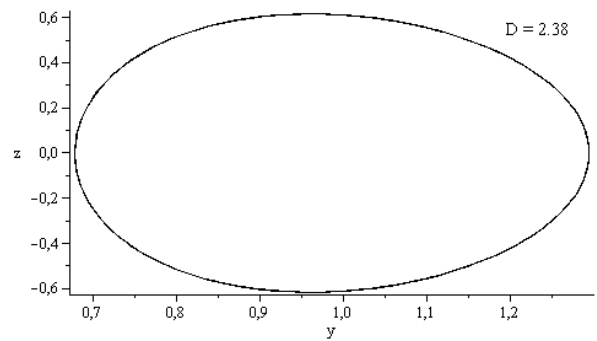}}} \caption{Phase trajectory in the plane
$y, z$ for periodic solution of period $\pi$, see central point in Fig. \ref{2-38}; $D=2.38$.} \label{2-38fd}
\end{figure}

The case $D=2.38$ is represented in Fig. \ref{2-38}. Layers with both regular and chaotic trajectories are
situated around periodic solutions. In Figs \ref{2-38ev}, \ref{2-38fd} time evolution $y(\tau)$, $z(\tau)$
and phase trajectory in the plane $y, z$ are shown for a periodic solution with period $\pi$.

\section{Transition to chaos via cascade of period doublings}

\begin{figure}
\centerline{\hbox{\includegraphics[width=0.4\textwidth]{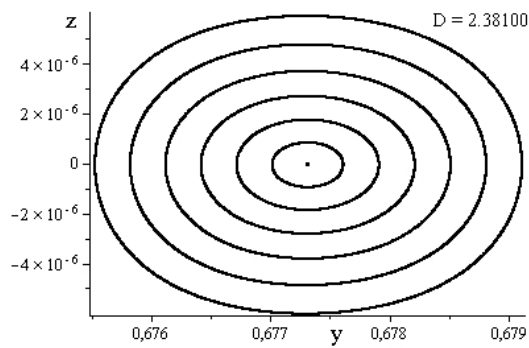}}} \caption{Poincar\'e section for $D=2.38100$.}
\label{2-38100}
\end{figure}

\begin{figure}
\centerline{\hbox{\includegraphics[width=0.4\textwidth]{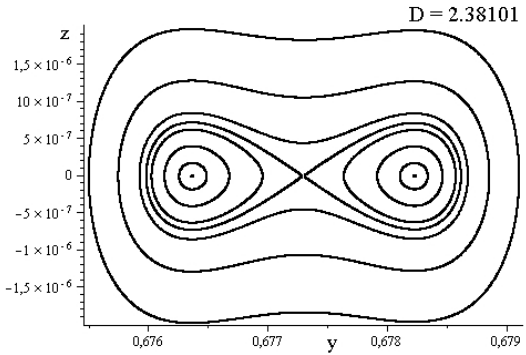}}} \caption{Poincar\'e section for
$D=2.38101$. This value of $D$ corresponds to first period doubling. The left central point maps to the right
central point after time $\pi$, then visa versa. Compare with Fig. \ref{2-38100}, where Poincar\'e section
just before this doubling is shown.} \label{2-38101}
\end{figure}

\begin{figure}
\centerline{\hbox{\includegraphics[width=0.465\textwidth]{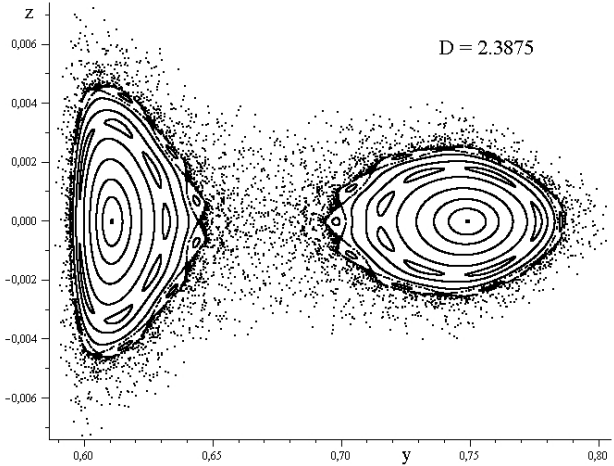}}} \caption{Poincar\'e section after
first period-doubling; $D=2.3875$. Two points in centres of concentric curves correspond to the periodic
solution of the period $2\pi$.} \label{2-3875}
\end{figure}

\begin{figure}
\centerline{\hbox{\includegraphics[width=0.45\textwidth]{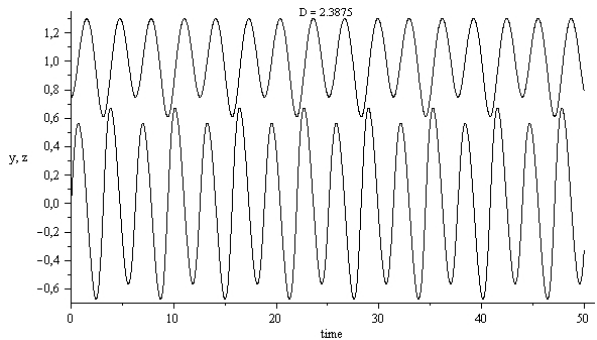}}} \caption{Time dependence of
non-dimensional radius $y$ (upper curve) and non-dimensional velocity $z = y'$ (lower curve) for the periodic
solution of period $2\pi$, see points in Fig. \ref{2-3875}; $D=2.3875$.} \label{2-3875ev}
\end{figure}

\begin{figure}
\centerline{\hbox{\includegraphics[width=0.45\textwidth]{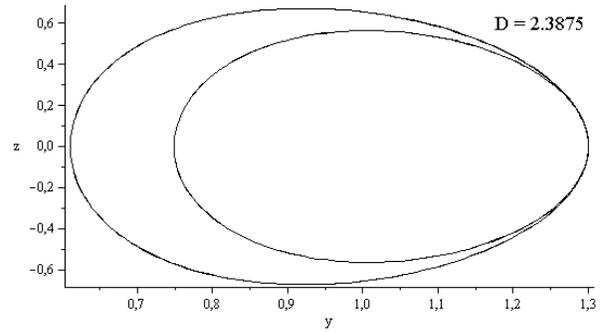}}} \caption{Phase trajectory in the
plane $y, z$ for periodic solution of period $2\pi$, see points in Fig. \ref{2-3875}; $D=2.3875$.}
\label{2-3875fd}
\end{figure}

\begin{figure}
\centerline{\hbox{\includegraphics[width=0.465\textwidth]{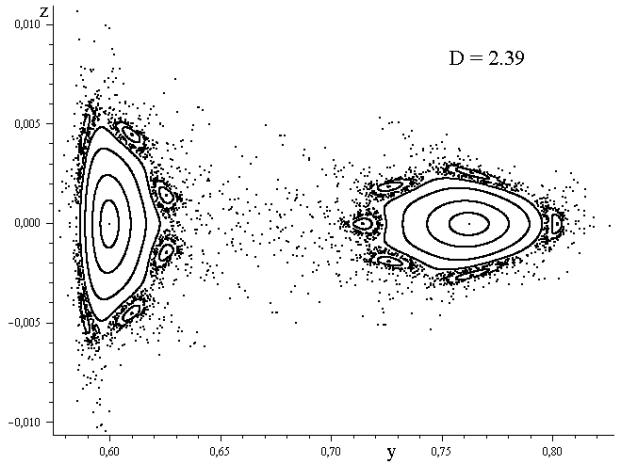}}} \caption{Poincar\'e section for
$D=2.39$. Two points in the centres of concentric curves correspond to {\bf a} periodic solution of period
$2\pi$.} \label{2-39}
\end{figure}

\begin{figure}
\centerline{\hbox{\includegraphics[width=0.45\textwidth]{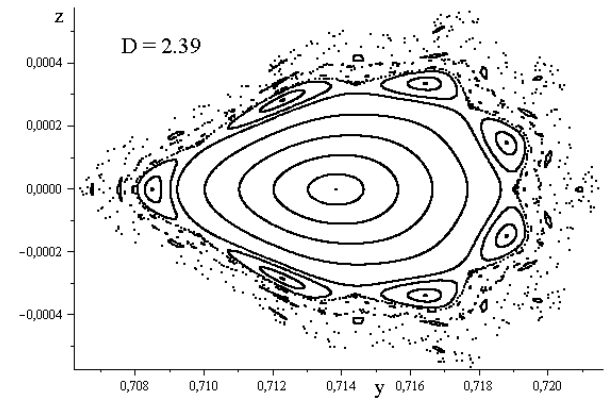}}} \caption{Zoom of left part of right
island in Fig. \ref{2-39}. The island at the Poincar\'e section for $D=2.39$. The point in the centre of the
island corresponds to the periodic solution of the period $12\pi$.} \label{2-39isla}
\end{figure}

With increasing $D$ the stable periodic solution considered in Sections 4 and 5 loses its stability via
period doubling bifurcation. The first period doubling occurs at $D_1=2.38101$. At this value $D_1$ the
solution with period $\pi$ becomes unstable, and a stable periodic solution of period $2\pi$ appears.

The subsequent points of period doubling $D_1$, $D_2$, ... can be found by constructing Poincar\'e sections
for different values of $D$. In Fig. \ref{2-38100} the central part of the Poincar\'e section for value of
$D$ before the first bifurcation is shown. In Fig. \ref{2-38101} the central part of the Poincar\'e section
for value $D$ soon after doubling is shown. Periodic solution with period $\pi$ is now unstable, but there is
a stable periodic solution of period $2 \pi$. On Poincar\'e section this solution is represented by two
points surrounded by closed curves (Fig. \ref{2-38101}). Under one iteration of Poincar\'e return map the
left point and surrounding it curves are mapped, respectively, onto right point and surrounding it curves,
and vice versa. With increasing $D$ the distance between the points of periodic solution on Poincar\'e
section increases.

In Fig.\ref{2-3875} the Poincar\'e section for a slightly greater value of $D$, namely $D=2.3875$ is shown.
In Figs \ref{2-3875ev}, \ref{2-3875fd} time evolution $y(\tau)$, $z(\tau)$ and phase trajectory in $y, z$
plane for the periodic solution with period $2\pi$ are shown.

At $D_2=2.410172$ the periodic solution with period $2\pi$ becomes unstable, and a stable periodic solution
of period $4\pi$ appears. The following doubling points found by the Poincar\'e section construction are:

\[ D_1 \simeq 2.38101, \quad \pi \rightarrow 2 \pi , \]
\[ D_2 \simeq 2.410172, \quad 2 \pi \rightarrow 4 \pi ,  \]
\[ D_3 \simeq 2.4137115, \quad 4 \pi \rightarrow 8 \pi ,  \]
\[ D_4 \simeq 2.4141168, \quad 8 \pi \rightarrow 16 \pi ,  \]
\[ D_5 \simeq 2.4141633, \quad 16 \pi \rightarrow 32 \pi .  \]

The values $D_1$, $D_2$, $D_3$, $D_4$ have also been obtained with the help of of AUTO-07P: Continuation and
Bifurcation Software for Ordinary Differential Equations (\citet{auto07}, http://cmvl.cs.concordia.ca/auto/).

According to \citet{feig81} the limiting constant

\begin{equation}
q = \frac{D_n-D_{n-1}}{D_{n+1}-D_n}
\end{equation}
should for large $n$ approach the constant $FH = 8.721…$ for Hamiltonian systems. We obtain the following values:
\[ q_{123} = \frac{D_2-D_1}{D_3-D_2} \simeq 8.24 , \]
\[ q_{234} = \frac{D_3-D_2}{D_4-D_3} \simeq 8.73 , \]
\[ q_{345} = \frac{D_4-D_3}{D_5-D_4} \simeq 8.72 . \]
We conclude that this behaviour agrees with the expected one for Hamiltonian systems, despite of the
additional symmetry $z \rightarrow -z$, $t \rightarrow -t$.

We can find approximately the value of parameter $D_{\infty}$ for which stable periodic solutions disappear
after an infinite cascade of period doublings using the value $q \simeq 8.721$:

\begin{eqnarray}
D_\infty &=& D_4 + (D_5-D_4) + (D_6-D_5) + (D_7-D_6) + ... \nonumber \\
&=& D_4 + (D_5 - D_4) + \frac{D_5-D_4}{q} + \frac{D_5-D_4}{q^2} + ... \nonumber \\
&=& D_4 + \frac{D_5-D_4}{1 - 1/q} \simeq 2.4141693.
\end{eqnarray}

In Table 1 we summarize the figures of the present article.

\begin{table}
\caption{Values of parameter $D$ and numbers of corresponding figures.}
\begin{tabular}{@{}cc} \hline
  Values of parameter $D$  & Numbers of figures  \\
  \hline
  1.5     &  \ref{phase1-5}             \\
  2.0     &  \ref{bk2-0}                \\
  2.001   &  \ref{2-001}, \ref{2-001big}, \ref{2-001fd}  \\
  2.1     &  \ref{bk2-1}                \\
  2.15    &  \ref{2-15}, \ref{2-15big}   \\
  2.25    &  \ref{bk2-25}, \ref{2-25}, \ref{2-25add}, \ref{2-25add2}    \\
  2.30    &  \ref{2-30}                      \\
  2.35    &  \ref{2-35}                          \\
  2.38    &  \ref{2-38}, \ref{2-38ev}, \ref{2-38fd}   \\
  2.38100 &  \ref{2-38100}                               \\
  2.38101 &  \ref{2-38101}                                   \\
  2.3875  &  \ref{2-3875}, \ref{2-3875ev}, \ref{2-3875fd}    \\
  2.39    &  \ref{2-39}, \ref{2-39isla}   \\
  2.4     &  \ref{bk2-4}                                  \\
  2.5     &  \ref{phase2-5}                               \\
  3.0     &  \ref{bk3-0}                                \\
  \hline
\end{tabular}

\end{table}

\section{Discussion}

In this section we would like to address the issue of a jet angular velocity oscillations generation. Jets
are associated with accretion and consist of ejected matter from accretion disc. Jet matter inherits angular
momentum from the accretion disc matter.

The jet matter may have a periodically (or quasi-periodically) varying sign of the angular momentum of the
accretion disc matter around a supermassive black hole. In a dense stellar cluster around the SMBH, the
accretion disc can change its direction of rotation due to different sign of the angular momentum of matter
of the disrupted accreting stars. In this situation the magneto-torsional oscillations should be inevitably
generated in the outflowing jets.

In the case, when stellar cluster is rotating, the disrupted stars will preserve the excess of the angular
momentum, and the jet may rotate, and there should be oscillations of the angular velocity around the regular
rotation. The presence of these oscillations will lead to effects similar to ones described in our model. But
the presence of regular rotation makes the physical picture of phenomenon more complicated, because of the
centrifugal force created by regular rotation and leading to jet expansion in radial direction.

In the framework of our model we have demonstrated that for a narrow range of the parameter $D$ the cylinder
radius remains finite for a long time. This can potentially lead to a long-live jets. Due to complicated
nature of the problem, we can not estimate the number of jets collimated by this mechanism. In the case of
$D<2$ the jet collimation does not occur. We can tell that if $D>2$ then jet exists in a continuous form or
in the form of blobs, both configurations corresponding to observations.

We can estimate value of $D$
\begin{equation}
D = \frac{1}{2\pi K C_m}\left(\frac{C_b \Omega_0}{z_0\omega}\right)^2
\end{equation}
in the following way:
\begin{equation}
\Omega_0 \simeq \frac{c}{R_0}, \quad z_0 \simeq R_0, \quad C_m \simeq \rho_0 R_0^2, \quad C_b \simeq B_{z,0}
R_0^2,
\end{equation}
\begin{equation}
\omega \simeq \alpha_n \frac{B_{z,0}}{z_0} \sqrt{\frac{\pi}{\rho_0}}, \quad K \simeq \frac{c^2}{A}.
\end{equation}
Here $c$ is the speed of light, $R_0$, $\rho_0$ and $B_{z,0}$ are some characteristic values of length,
density and magnetic field correspondingly, $A$ is the non-dimensional constant. We obtain:
\begin{equation}
D = \frac{A}{2 \pi^2 \alpha_n^2} .
\end{equation}
For estimation of $A$ we can write: $P = nkT$, $n=\frac{\rho_0}{m_p}$, where $P$ is the pressure, $n$ is the
concentration, $T$ is the temperature, $m_p$ is the proton mass, $k$ is the Boltzmann constant. Comparing
with $P=K\rho_0$ we obtain:
\begin{equation}
A = \frac{m_p c^2}{kT}.
\end{equation}
For estimation in a nonlinear regime of oscillations we can put $\alpha_n^2=0.1$ (see \cite{bk07}).

We see that for a sufficiently large proton loading ($A>1$) we can have $D > 2$.

\section*{Acknowledgments}
The work of GSBK, OYuT and YuMK was partially supported by the Russian Foundation for Basic Research grants
08-02-00491 and 11-02-00602, the RAN Program 'Origin, formation and evolution of objects of Universe' and
Russian Federation President Grant for Support of Leading Scientific Schools NSh-3458.2010.2.

The work of AIN was partially supported by the Russian Foundation for Basic Research grant 09-01-00333, and
the President of the Russian Federation Grant for Support of Leading Scientific Schools NSh-8784.2010.1.

The work of OYuT was partially supported by the Dynasty Foundation.

The work of OYuT and YuMK was also partially supported by Russian Federation President Grant for Support of
Young Scientists MK-8696.2010.2.

OYuT is grateful to O.D. Toropina for help in the artwork preparation.

\appendix

\section[]{Approximate formulas for periodic solutions and invariant curves}

We will use a standard approach for the averaging method. We will make a canonical time-periodic
transformation of variables such that the Hamiltonian for the new variables will not depend on time in the
principal approximation. Discarding small time-depending terms in the Hamiltonian we will get a Hamiltonian
for some autonomous system. We will find an equilibrium position of this system. The exact system in the new
variables has a periodic solution close to this equilibrium position. Then we will find formulas for this
periodic solution in the old variables making use of the formulas for the transformation of variables.

We will construct the required transformation of variables as a composition of several transformations of
variables. In the system with Hamiltonian (\ref{H_nonav}) we make a canonical transformation of variables
$(q,z)\mapsto(\bar q,\bar z)$ with a generating function of the form

\begin{equation}
q\bar z+ \varepsilon S_1(q,\bar z ,\tau).
\end{equation}
The old variables $(q,z)$ and the new variables $(\bar q,\bar z)$ are related via formulas

\begin{equation}
z=\bar z+ \varepsilon \frac{\partial S_1}{\partial q},\quad  \bar q=q+ \varepsilon \frac{\partial S_1}{\partial \bar
z}.
\end{equation}
The dynamics of the new variables is described by the Hamiltonian

\begin{eqnarray}
\varepsilon \bar H &=& \varepsilon H+\varepsilon \frac{\partial S_1}{\partial \tau}= \varepsilon \left[
\frac{1}{2}\left(\bar z + \varepsilon \frac{\partial S_1}{\partial q}\right )^2 \right. - \nonumber \\
&-& \left. \left(1-\frac{D}{2}\left(1-\cos 2\tau\right)\right )\ln q+ \frac{\partial S_1}{\partial \tau} \right].
\end{eqnarray}
Let us choose $S_1$ such that the terms of order $\varepsilon$ in this Hamiltonian will be independent on
$\tau$:

\begin{equation}
-\frac{D}{2} \ln q \cos 2\tau+ \frac{\partial S_1}{\partial \tau}=0
\end{equation}
One can take

\begin{equation}
S_1=\frac{D}{4}\ln q \sin 2\tau.
\end{equation}
Then

\begin{equation}
z=\bar z +\varepsilon\frac{D}{4q}\sin 2\tau,\quad \bar q=q.
\end{equation}
The new Hamiltonian is

\begin{eqnarray}
\varepsilon \bar H &=& \varepsilon \left [ \frac{1}{2}\left(\bar z + \varepsilon\frac{D}{4\bar q}\sin 2\tau\right )^2+
\frac{1}{2}(D-2)\ln \bar q\right ] \nonumber \\
&=& \varepsilon \left [ \frac{1}{2}\bar z^2+\varepsilon\frac{D\bar z}{4\bar q}\sin 2\tau
+\varepsilon^2\frac{D^2}{32\bar q^2}\sin^2 2\tau \right. \nonumber \\
&+& \left. \frac{1}{2}(D-2)\ln \bar q \right ].
\end{eqnarray}
Now we make a canonical transformation of variables $(\bar q,\bar z)\mapsto(\tilde q,\tilde z)$ in the system
with Hamiltonian $\varepsilon \bar H$ with the generating function of the form

\begin{eqnarray}
\bar q\tilde  z+ \varepsilon^2 S_2(\bar q,\tilde z ,\tau) .
\end{eqnarray}
The old variables $(\bar q,\bar z)$ and the new variables  $(\tilde q,\tilde z)$ are related via formulas

\begin{eqnarray}
\bar z=\tilde z+ \varepsilon^2 \frac{\partial S_2}{\partial \bar q},\quad \tilde q=\bar q+ \varepsilon^2 \frac{\partial
S_2}{\partial \tilde z}.
\end{eqnarray}
The dynamics of the new variables is described by the Hamiltonian

\begin{eqnarray}
\varepsilon  \bar{\bar H}&=& \varepsilon \bar H  +  \varepsilon^2 \frac{\partial S_2}{\partial \tau}\nonumber
\\&=&
\varepsilon \left [ \frac{1}{2}\left(\tilde z  +  \varepsilon^2   \frac{\partial S_2}{\partial \bar  q}\right )^2
+\varepsilon\frac{D}{4\bar q}\left(\tilde z  +  \varepsilon^2 \frac{\partial S_2}{\partial \bar  q}\right )\sin
2\tau \right. \nonumber
\\&+&
\left. \varepsilon^2\frac{D^2}{32\bar q^2}\sin^2 2\tau+ \frac{1}{2}(D-2)\ln \bar q
 +\varepsilon \frac{\partial S_2}{\partial \tau}
\right ].\nonumber
\end{eqnarray}
Let us choose $S_2$ in such a form that the terms of order $\varepsilon^2$ in this Hamiltonian will be
independent on $\tau$:

\begin{eqnarray}
\frac{D\tilde z}{4\bar q}\sin 2\tau+
 \frac{\partial S_2}{\partial \tau}=0.
\end{eqnarray}
One can take

\begin{eqnarray}
S_2=\frac{D\tilde z}{8\bar q}\cos 2\tau.
\end{eqnarray}
Then

\begin{eqnarray}
\bar z=\tilde z-\varepsilon^2\frac{D\tilde z}{8\bar q^2}\cos 2\tau, \quad \tilde q=\bar q+\varepsilon^2\frac{D}{8\bar
q}\cos 2\tau .
\end{eqnarray}
The new Hamiltonian is

\begin{eqnarray}
\varepsilon\bar{\bar H}&=& \varepsilon\left [\frac {1}{2}(\tilde z-\varepsilon^2\frac{D\tilde z}{8\bar q^2}\cos
2\tau)^2-\varepsilon^3\frac{D^2\tilde z}{32\bar q^3}\cos 2\tau\sin 2\tau \right. \nonumber
 \\&+&
 \left. \frac{\varepsilon^2D^2}{32\bar
q^2}\frac{1-\cos2\tau}{2}+\frac{1}{2}(D-2)\ln\bar q\right].
\end{eqnarray}


In this relation one should express $\bar q$ through $\tilde q$. Let us suppose that $(D-2)\sim
\varepsilon^2$ (only for such values of $D$ the periodic solution will exist). Then we will just replace
$\bar q$ with $\tilde q$; this will lead to the error $O(\varepsilon^5)$ in the Hamiltonian. Thus we have

\begin{eqnarray}
\varepsilon\bar{\bar H}&=& \varepsilon\left [\frac {1}{2}\tilde z^2-\varepsilon^2\frac{D\tilde z^2} {8\tilde q^2}\cos
2\tau+\frac{\varepsilon^2D^2}{32\tilde q^2}\frac{1-\cos2\tau}{2} \right. \nonumber
\\&+&
\left.  \frac{1}{2}(D-2)\ln\tilde q\right] + O(\varepsilon^4).
\end{eqnarray}
Now we make a canonical, $O(\varepsilon^3)$-close to the identical transformation of variables $(\tilde
q,\tilde z)\mapsto(\hat q,\hat z)$ in the system with Hamiltonian $\varepsilon \bar {\bar H}$ such that the
terms of order $\varepsilon^3$ in the new Hamiltonian will be independent on $\tau$. The new Hamiltonian in
the principal approximation is just the average of the old Hamiltonian over $\tau$. Thus the new Hamiltonian
is

\begin{equation}
\varepsilon\bar{\bar{\bar H}}=\varepsilon\left [\frac{1}{2}\hat z^2+\frac{\varepsilon^2D^2} {64\hat q^2}+\frac{1}{2}
(D-2)\ln\hat q\right ]+O(\varepsilon^4).
\end{equation}
Let us neglect the term $O(\varepsilon^4)$ in this Hamiltonian. We obtain an autonomous Hamiltonian system.
Let us demonstrate that this system has an equilibrium position. We should calculate partial derivatives of
the Hamiltonian and find a point where they vanish. Thus for the coordinates of the equilibrium position we
have

\begin{equation}
\hat z=0,\quad -\frac{\varepsilon^2D^2}{32\hat q^3}+\frac{1}{2\hat q}(D-2)=0.
\end{equation}
Thus we have

\begin{equation}
\hat q^2=\frac{\varepsilon^2D^2}{16(D-2)}.
\end{equation}

In the numerator of this formula we can put $D=2$ as this will lead to an error $O(\varepsilon^2)$ in $q$.
Thus for an approximate position $\hat q_*$ of the equilibrium we have
\begin{eqnarray}
\hat q_*=\frac{\varepsilon }{2\sqrt {D-2}}.
\end{eqnarray}
Plugging into previous formulas for transformation of variables the coordinate of the equilibrium $(\hat q,
\hat z)= (\hat q_*, 0)$ we find a periodic solution of the original system. Approximate formulas for this
solution are

\begin{eqnarray}
q &\approx& \frac{\varepsilon }{2\sqrt {D-2}}-\varepsilon^2 \frac {D}{8q_*}\cos{2\tau}\approx \frac{\varepsilon}{2\sqrt
{D-2}} -\varepsilon^2 \frac{2\cos 2\tau}{\frac{8\varepsilon}{2\sqrt{D-2}}} = \nonumber \\
&=& \frac{\varepsilon}{2\sqrt {D-2}}-\frac{\varepsilon}{2}\sqrt {D-2}\cos 2\tau,\nonumber \\
z &\approx& \varepsilon\frac{D}{4q_*}\sin 2\tau = \varepsilon\frac{2\sin 2\tau}{4\varepsilon /{2}\sqrt{D-2}}=
\sqrt{D-2}\sin 2\tau.\nonumber
\end{eqnarray}
Returning to $y=q/\varepsilon$ we get

\begin{equation}
y=\frac{1}{2\sqrt{D-2}}-\frac{1}{2}\sqrt{D-2}\cos2\tau, \;\; z=\sqrt{D-2}\sin 2\tau.
\end{equation}
The function $\bar{\bar{\bar H}}$ is an approximate first integral of motion: $\bar{\bar{\bar H}}\approx {\rm
const}$. From this relation we obtain an approximate expression for the invariant curves of the Poincar\'e
return map:

\begin{equation}
\frac{1}{2}z^2+ \frac{1}{16y^2}+\frac{1}{2}(D-2)\ln y = {\rm const}.
\end{equation}

\label{lastpage}

\end{document}